\documentclass[twocolumn,showpacs,amsmath,aps]{revtex4}

\usepackage{graphicx}
\usepackage{dcolumn}
\usepackage{bm}

\begin{document}

\title{Laser induced birefringence in a wavelength-mismatched cascade system of inhomogeneously broadened Yb atoms}

\author{Tai Hyun Yoon}
\email{thyoon@kriss.re.kr}
\affiliation{Center for Optical Frequency Control\\ Korea Research Institute of Standards and Sciecne\\
1 Doryong, Yuseong, Daejeon 305-340, Korea}
\author{Chang Yong Park}
\affiliation{Center for Optical Frequency Control\\ Korea Research Institute of Standards and Sciecne\\
1 Doryong, Yuseong, Daejeon 305-340, Korea}
\author{Sung Jong Park}
\affiliation{Center for Optical Frequency Control\\ Korea Research Institute of Standards and Sciecne\\
1 Doryong, Yuseong, Daejeon 305-340, Korea}

\date{July 22, 2004}

\begin{abstract}
We report the observation of laser induced birefringence (LIB) in a wavelength-mismatched cascade
system ($J=0 \leftrightarrow J=1 \leftrightarrow J=0$ transitions) of inhomogeneously broadened
ytterbium atoms with strong pump and probe fields. We investigate the transmission spectrum of two
circular polarization ($\sigma_p^+$ and $\sigma_p^-$) components of strong probe field at fixed
frequency, depending on the detuning of circularly polarized ($\sigma_c^-$) coupling field from
two-photon resonance. We find that $\sigma_p^+$ ($\sigma_p^-$) polarized component exhibits a
narrow electromagnetically induced absorption (transparency) spectrum. Numerical solutions of
density matrix equations show qualitative agreement with experimental results. A Doppler-free
dispersive LIB signal is obtained by detecting the Stokes parameter of the probe field, enabling us
to stabilize the frequency of coupling laser without frequency modulation.
\end{abstract}

\maketitle

Quantum coherence effects in atomic three-level systems interacting with two monochromatic laser
fields provide a rich source of interesting phenomena in laser spectroscopy, for instance, the
phenomena of electromagnetically induced transparency (EIT) and absorption (EIA) \cite{HA}, etc.
Complete coherent control of the polarization state of an optical field \cite{Wielandy},
measurement of the optical rotation to investigate the atomic parity nonconservation in thallium
vapor \cite{Cronin}, and measurement of birefringence at 1083 nm in $^4$He atoms \cite{Pavone} have
been achieved relying on the EIT effects in the corresponding cascade three-level atomic systems.
Recently, Magno and coworkers \cite{Magno} proposed a simple scheme for laser cooling of
alkaline-earth and ytterbium atoms using a two-photon $^1S_0\leftrightarrow$ $^1P_1\leftrightarrow$
$^1S_0$ cascade transition, which might be used as a second stage cooling after precooling with the
$^1S_0 - $ $^1P_1$ transition, and this two-photon laser cooling scheme has not been realized yet
experimentally.

In this paper, we present a Doppler-free two-photon atomic coherence spectroscopy of Yb atoms
associated with the wavelength-mismatched cascade scheme considered in the two-photon cooling of Yb
atoms \cite{Magno}. We investigate the Doppler-free transmission spectrum of two circular
polarization ($\sigma_p^+$ and $\sigma_p^-$) components of strong probe field at fixed detuning
from one-photon resonance in a Yb hollow cathode lamp (HCL) depending on the detuning of circularly
polarized ($\sigma_c^-$) counterpropagating coupling field from two-photon resonance. We find that
when the Rabi frequencies of the probe and coupling fields are similar to the decay rate of the
$J=1$ state, $\sigma_p^+$ component exhibits a narrow electromagnetically induced absorption (EIA)
spectrum, while $\sigma_p^-$ component exhibits an electromagnetically induced transparency (EIT)
spectrum due the multiphoton process induced via the strong probe field. The measured spectra show
qualitative agreement with the calculated spectra from numerical solutions of semiclassical density
matrix equations. A sensitive Doppler-free dispersive signal which is proportional to the
birefringence experienced by two circular polarization components of the probe field is then
subsequently obtained by detecting the Stokes parameter \cite{Pearman} of the transmitted probe
field. This is the first measurement, as far as the authors know, of the electromagnetically
induced birefringence originated from the EIA-type resonance of the strong probe field in the
wavelength-mismatched three-level cascade system. Finally, we are able to stabilize the frequency
of coupling laser at 1077 nm by use of the laser induced birefringence (LIB) signal with long-term
stability below 1 MHz, which is stable enough to study the two-photon laser cooling of Yb atoms in
a Yb magneto-optical trap (MOT) \cite{Park2}.

\begin{figure}
\includegraphics[height=8cm,width=3.3cm,angle=-90]{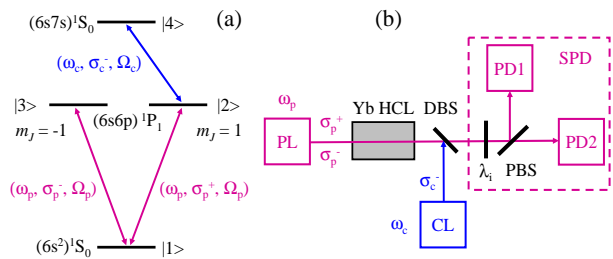}
\caption{(a) The three-level cascade scheme of Yb atoms having $m$-degenerate sublevels $|2\rangle$
and $|3\rangle$. In parenthesis, laser frequency, polarization state, and Rabi frequency of the
driving fields are explicitly shown. (b) Experimental set-up for Doppler-free two-photon atomic
coherence spectroscopy. PL; probe laser, CL; coupling laser, SPD; Stokes parameter detector, PD;
photo-diode, PBS; polarization beam splitter, DBS; dichroic beam splitter, HCL; hollow cathode
lamp, $\lambda_i$; $ \lambda/i$ wave-plate ($i = 2, 4$). \label{fig1}}
\end{figure}

Figure~\ref{fig1}(a) and (b), respectively, show the three-level cascade scheme of Yb atoms having
$m$-degenerate sublevels $|2\rangle$ and $|3\rangle$ relevant to the present study and the
schematic diagram of the experimental set-up. In Fig.~\ref{fig1}(a), the levels $|1\rangle$,
$|2\rangle$, and $|4\rangle$ form a three-level cascade wavelength-mismatched configuration and the
level $|2\rangle$ and $|3\rangle$ are coupled to the ground state $|1\rangle$ by an arbitrarily
intense probe field. Here we ignore the state with $m_J=0$ of the $^1P_1$ state for simplicity, but
include it for the numerical calculation (see later) \cite{Patnaik}. In parenthesis on
Fig.~\ref{fig1}(a), laser frequency, polarization state, and Rabi frequency of the driving fields
are shown explicitly. The $^1P_1$ state (levels $|2\rangle$ and $|3\rangle$) decays rapidly with
decay rate $\gamma_a = 2\pi \times 28$ MHz to the level $|1\rangle$ and there is a weak decay
channel to the $(6s6p)^3P_1$ state via the levels $(6s5d)^3D_1$ and $^3D_2$ (not shown), and the
$^3P_1$ state decays to the ground state by emitting 556 nm spontaneous photons with decay rate
$2\pi \times 182$ kHz. The level $|4\rangle$ decays predominantly to the $^1P_1$ state with decay
rate $\gamma_b=2\pi\times 3.0$ MHz and weakly to the $^3P_1$ state with decay rate of $ 2\pi \times
0.2$ MHz \cite{Bai}.

Note that in Yb cascade three-level configuration in Fig.~\ref{fig1}(a), where $\omega_p \sim
2.7\times\omega_c$, the effects of mismatching wavelength in the EIT spectrum in Doppler-broadened
media is expected to be very strong \cite{Boon}, i.e., the depth of EIT dip is greatly reduced from
the complete transparency as opposed to the case when $\omega_p \simeq \omega_c$ \cite{Wielandy2}
and $\omega_c \sim 1.5 \times \omega_p$ \cite{Pavone}. In this case, the Autler-Townes components
for nonzero velocity atoms are completely overlap with line center as oppose to the case when
$\omega_p \leq \omega_c$, thus one would reasonably expect transparency created on resonance to be
obscured \cite{Boon}. This is exactly what happens in the cascade system of Fig.~\ref{fig1}(a) and
observed in Fig.~\ref{fig2} below, where we could detect only a few percent contrast of EIT or EIA
signals against Doppler profile only when the Rabi frequencies of probe and coupling fields are an
order of the decay rate of the $J=1$ state.

Theoretical analysis of the LIB in the same level configuration as in Fig.~\ref{fig1} was discussed
previously by Patnaik and Agarwal \cite{Patnaik} at the {\it{weak}} probe-field limit and
wavelength {\it{matched}}-configuration, i.e., $\omega_p = \omega_c$. We find, however, that in the
wavelength-{\it{mismatched}} and inhomogeneously broadened cascade configuration, where $\omega_p
\gg \omega_c$ as in Ca, Sr, and Yb atoms (considered in the two-photon cooling theory
\cite{Magno}), the LIB signal becomes negligibly small after Doppler averaging \cite{Boon} when the
probe field is weak. Thus, we can not use the analytical results discussed in Ref.~\cite{Patnaik}
for the explanation of our experimental findings. In this paper, we demonstrate that if the probe
field is {\it{strong}} in this level configuration two circular polarization components of the
probe field exhibit EIA in the $\sigma_p^+$ component which is coupled to the control field and EIT
in the $\sigma_p^-$ component which is decoupled from the control field, allowing us to measure the
laser induced birefringence originated from the EIA-type resonance. This is the first observation,
as far as the authors know, of an electromagnetically induced birefringence associated with the EIA
and EIT effects of the strong probe field in the cascade three-level system.

In the experiment, two laser diodes with external grating feedback are used for the probe laser
(PL) and coupling laser (CL), respectively, as indicated in Fig.~\ref{fig1}(b). The atomic
temperature in a Yb HCL is estimated to be $\sim$ 600 K, which gives a Doppler width of $\sim$ 1
GHz \cite{Park1}.  The frequency of probe laser at 399 nm can be stabilized within the range of
probe detuning $|\delta_p| < 2\gamma_a \sim 60$ MHz by use of a Doppler-free saturation absorption
signal with lock-in detection using a separate Yb HCL. Two-photon Doppler cancellation, although it
is not perfect since $\omega_p\ne\omega_c$, is achieved with counterpropagating geometry of the
coupling and probe fields as in Fig.~\ref{fig1}(b). We set the powers of the probe and coupling
fields at $P_{p}$ = 85 $\mu$W (spot size 67$\mu$m$, \Omega_p \sim 2 \gamma_a$ for $\sigma_p^+$ and
$\sigma_p^-$ components) and $P_{c}$ = 1.7 mW (spot size 175 $\mu$m, $\Omega_c \sim 4 \gamma_a$),
and the probe field detuning at exact one-photon resonance ($\delta_p = 0$).

We first measured the transmittance spectrum corresponding to the two circularly polarized
components of the probe field as a function of coupling field detuning $\delta_c/\gamma_a$. For
this measurement, we used two detectors (PD1 and PD2) in Fig.~\ref{fig1}(b) and a $\lambda/4$
wave-plate oriented at 45$^{\text{ o}}$ to the PBS. The first column in Fig.~\ref{fig2} shows the
typical transmission spectrum expressed as a contrast against Doppler background (70 \% absorption)
corresponding to the $\sigma_p^-$ component in (a) and to the $\sigma_p^+$ component in (b) for the
$^{174}$Yb isotope in a Yb HCL with Ne buffer gas (discharge voltage 160 V and current 1.8 mA)
\cite{Park2}. As one can see, the spectrum in (a) corresponds to the $\sigma_p^-$ component shows
an EIT spectrum with a line width of $\sim3\gamma_a$, while the spectrum in (b) corresponding to
the $\sigma_p^+$ component shows a narrower ($\sim\gamma_a$) EIA spectrum at the two-photon
resonance. Note that the EIT and EIA spectrum has only a few percent contrast against Doppler
background, which is expected theoretically for the wavelength-mismatched cascade system with
$\omega_p\gg\omega_c$ \cite{Boon}. We also find that, as decreasing the intensity of coupling
field, the EIA spectrum in (b) develops into the EIT spectrum as in (a) with reduced
signal-to-noise ratio. However, the $\sigma_p^-$ component which is decoupled from the strong
control field exhibits always an EIT spectrum under the parameter range we have tested. The EIT
spectrum has a line width about 87 MHz $\sim 3\gamma_a$, which is 1.5 times larger than the width
measured at the Doppler-free saturation absorption spectroscopy with a same Yb HCL due to the
saturation broadening \cite{Park2}, while the EIA spectrum has much narrow line width about
$\gamma_a$. We calibrated the width of the measured spectrum from the known frequency interval
(isotope shift) of the $^1S_0$ $ - $ $^1P_1$ transition of Yb isotopes \cite{Loftus}. The narrow
feature of the EIA spectrum may be understood that in the EIA process higher order multi-path
interference is involved compared to the EIT process, since the probe field is strong
\cite{Wielandy2}.

In order to understand theoretically the measured spectra in Fig.~\ref{fig2}(a) and (b), we solved
a semiclassical density matrix equation associated with the level configuration of
Fig.~\ref{fig1}(a) at the steady state. The equation of motion for the slowly varying components
$\rho$ of the density matrix $\tilde{\rho}$ may be written by making a transformation
$\tilde{\rho}\rightarrow\rho$ such that $\rho_{kk} = \tilde{\rho}_{kk}$, $\rho_{l1} =
\tilde{\rho}_{l1} \exp(-i\omega_p t)$, $\rho_{4l} = \tilde{\rho}_{4l}\exp(-i\omega_c t)$,
$l=2,o,3$, $\rho_{41} = \tilde{\rho}_{41}\exp[-i(\omega_p+\omega_c)t]$, where subscript $"o"$
indicates the state with $m=0$ of the $^1P_1$ level. The matrix equation for $\rho$ is then found
to be \cite{Patnaik}
\begin{eqnarray}\label{eqr1}
\dot{\rho} = -\frac{i}{\hbar}[\it{H},\rho]&-&\sum_{i=2,o,3}(\gamma_b/6
\{|\rm{4}\rangle\langle\rm{4}|,\rho\}+\gamma_a/2
\{|{\rm{i}}\rangle\langle\rm{i}|,\rho\}\nonumber\\
&-&\gamma_b\rho_{44}|\rm{i}\rangle\langle\rm{i}|-\gamma_a\rho_{ii}
|\rm{1}\rangle\langle\rm{1}|),
\end{eqnarray}
with the effective hamiltonian in the rotating frame
\begin{eqnarray}\label{eqr2}
H &=& \hbar(\delta_p+\delta_c)|4 \rangle \langle 4| + \hbar\delta_p \sum_{i=2,3}
|{\rm{i}}\rangle\langle {\rm{i}}|\nonumber \\
&&-\frac{\hbar}{2} \sum_{i=2,3}(\Omega_p|{\rm{i}}\rangle\langle1|+ \Omega_c|4\rangle\langle
{\rm{2}}|+ {\rm{h.c.}}).
\end{eqnarray}
The second term under the summation sign of Eq.~(\ref{eqr1}) represents the natural decays of the
system and the curly bracket represents the anti-commutator.

The second column in Fig.~\ref{fig2} ((c) and (d)) shows the calculated transmission spectrum of
the probe field after Doppler averaging expressed as a contrast against Doppler background as a
function of $\delta_c/\gamma_a$ with the parameters $\Omega_p = \Omega_c = 2\gamma_a$. The
calculated contrast of the probe field transmittance against Doppler background was obtained from
the imaginary part of the atomic coherence, i.e., Im[$\rho_{31}$] for $\sigma_p^-$ component and
Im[$\rho_{21}$] for $\sigma_p^+$ component. The spectrum in (c) shows an EIT spectrum as in (a),
while the spectrum in (d) corresponding to the $\sigma_p^+$ component shows an EIA spectrum as in
(b) at the same parameters. The EIA spectrum of the $\sigma_p^+$ component of the probe field can
be understood as a result of higher-order multiphoton interference enabled by the strong probe
field \cite{Wielandy2}. In addition, our experimental and numerical observations suggest that the
$\sigma_p^-$ component which is blind to the coupling field is also actually strongly coupled to
the $\sigma_c^-$ coupling field due to the multiphoton process induced via the strong probe field
so that the EIT spectrum is observed.

Although, as one can find, the experimental and theoretical spectra in Fig.~\ref{fig2} agree well
qualitatively, there are significant difference in line shape. We attribute this difference partly
to the unknown discharge effects in the Yb hollow-cathode lamp and partly to the uncontrollable
imperfections of the polarization elements in the experimental set-up. However, we can reasonably
say that the density matrix equations in Eqs.~(\ref{eqr1}) and (\ref{eqr2}) correctly describes the
general behavior of the atomic coherence associated with the three-level cascade system with
$m$-degenerate sublevels in Fig.~\ref{fig1}(a). We want to emphasize that at the parameter range
where $\Omega_p \ll \gamma_a$ and $\Omega_c \sim \gamma_a$, the calculated transmittances for both
$\sigma_p^+$ and $\sigma_p^-$ components exhibit white (flat) spectrum after the Doppler averaging,
thus there is no detectable EIT or EIA signals as contrary to the case when $\Omega_p \sim
\Omega_c\sim\gamma_a$ as in Fig.~\ref{fig2}, supporting strongly our observations. Therefore it is
very important experimental and theoretical observation that only when the intensity of the probe
field and coupling fields are strong, i.e., $\Omega_p \sim \Omega_c \sim \gamma_a$, the
transmittances of the $\sigma_p^+$ and $\sigma_p^-$ components in the wavelength-mismatched cascade
system exhibit an enhanced absorption or transmittance induced by the narrow two-photon coherence,
resulting in the enhancement of laser induced birefringence as clearly demonstrated in
Fig.~\ref{fig3}(a) below.

Strong coupling and probe fields, when applied to an initially isotropic medium containing Yb atoms
having $m$-degenerate sublevels, can create birefringence in the medium. Because, as clearly seen
in Fig.~\ref{fig2}, the strong driving fields create asymmetry between the susceptibilities
$\chi^{\pm}$ of the probe field, since $\chi^{+(-)} = (2\mu/\epsilon_o E_p)~\rho_{21(31)}$, where
$\mu$ is the dipole matrix element between $^1S_0$ - $^1P_1$ transition, $\epsilon_0$ is the
permittivity, and $E_p$ is the amplitude of the probe field. That results in the LIB, i.e, the
plane of polarization is rotated when it passes through the medium. For a small absorption the
rotation angle $\Delta\theta$ is given by \cite{Pearman,Patnaik}
\begin{equation}\label{eqr3}
\Delta\theta = \pi k_p L(\chi^+-\chi^-)=\Delta n k_pL/2,
\end{equation}
where $k_p$ corresponds to propagation vector of the probe field, $L$ is length of the cell along
$k_p$, $\Delta n = n^+-n^-$, $n^{\pm}+i\alpha^{\pm}=\sqrt{1+4\pi\chi^{\pm}}$, and $n^{\pm}$
($\alpha^{\pm}$) being the refractive index (absorption coefficient). Using the numerical solutions
for $\chi_p^{\pm}$ from Eqs.~(\ref{eqr1}) and (\ref{eqr2}), the rotation of polarization
$\Delta\theta$ of the probe can easily be determined from Eq.~(\ref{eqr3}). In order to detect the
background-free dispersive LIB signal experimentally, we used the SPD in Fig.~\ref{fig1}(b). In
this case, the SPD consists of a $\lambda/2$-wave plate and a balanced polarimeter \cite{Pearman}.
The conventional way to measure the polarization state after passing through the medium is to put a
linear polarizer in front of a detector and measure the projected intensity. But, in that case, it
is inevitable to include the effect of circular dichroism in addition to the rotation of the
polarization axis. In the SPD scheme, however, the $\lambda/2$-wave plate rotates the polarization
axis of the probe beam by 45$^{~\rm{o}}$, and the balanced detector (polarimeter) which subtracts
the reflected and transmitted intensities from the PBS measures only the rotation of the
polarization axis, i.e., the LIB signal which is linear in $\Delta\theta$ without background as
long as $\Delta\theta \ll 1$ for optically thin sample, where $\alpha L \ll 1$ \cite{Pearman}.

Figure~\ref{fig3}(a) shows the detected LIB signal at the same parameter condition as in
Fig.~\ref{fig2}. The LIB signal has two distinct features originated from the absorption
(dispersion) spectrum of the $\sigma_p^+$ and $\sigma_p^-$ components. The narrow dispersion
feature at the line center with line width about $\sim\gamma_a$ corresponds to the LIB associated
with the EIA signal observed for the $\sigma_p^+$ component in Fig.~\ref{fig2}(b), while somewhat
broad ($\sim3\gamma_a$) and weak dispersion signal superimposed on the main LIB signal corresponds
to the LIB associated with the EIT signal observed for the $\sigma_p^-$ component in
Fig.~\ref{fig2}(a). Therefore, we demonstrated for the first time the measurement of
electromagnetically induced birefringence originated from the EIA-type resonance of the strong
probe field in the wavelength-mismatched three-level cascade system.

Finally, we used successfully the measured dispersive LIB signal in Fig.~\ref{fig3}(a) for the
frequency stabilization of the coupling laser without frequency modulation with long-term stability
below 1 MHz as shown in Fig.~\ref{fig3}(b). Furthermore, by stabilizing the frequency of probe
laser to the different isotope of Yb atoms, for example $^{171}$Yb, which is a promising candidate
for future optical lattice clock \cite{Park2,Porsev}, we were able to stabilize the frequency of
coupling laser to each stable Yb isotope. The two-photon cooling theory predicts that the minimum
temperature of 124 $\mu$K for Yb atoms can be reached at the probe and coupling laser detunings of
$\delta_p=-\gamma_a/2$ and $\delta_c=-\gamma_b/2$ \cite{Magno}, respectively, those conditions can
easily be achieved with the current frequency-stabilized laser diodes described in this paper.

In summary, we introduced a Doppler-free two-photon atomic coherence spectroscopy of Yb atoms in a
Yb hollow cathode lamp associated with the wavelength-mismatched cascade level configuration. We
investigated the transmission spectrum of strong probe field at fixed one-photon frequency,
depending on the detuning of circularly polarized coupling field from two-photon resonance. We
found that $\sigma_p^+$ polarized component which is coupled to the upper state exhibits a narrow
EIA spectrum, while $\sigma_p^-$ polarized component exhibits an EIT spectrum. Numerical solutions
of density matrix equations show qualitative agreement with the experimental results. A
Doppler-free dispersive LIB signal results from the EIA-type resonance was obtained for the first
time by detecting the Stokes parameter of the probe field, enabling us to stabilize the frequency
of coupling laser without frequency modulation. The Doppler-free two-photon atomic coherence
spectroscopy introduced in this paper might equally be applied to the alkaline-earth atoms and
should be useful for experimental realization of the two-photon cooling theory \cite{Magno}.

This research is supported by the Creative Research Initiatives Program of the Ministry of Science
and Technology of Korea. The authors thank H. S. Moon for useful discussions.

\vspace{2cm}

\newpage

\begin{figure}
\includegraphics[height=8cm,width=6cm,angle=-90]{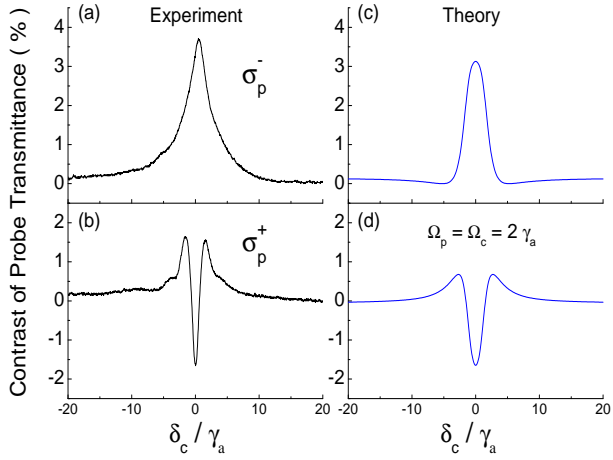}
\caption{Measured contrast against Doppler background of the strong probe field transmittance for
$\sigma_p^-$ (a) and $\sigma_p^+$ component (b) versus $\delta_c/\gamma_a$ at probe and coupling
powers of 85 $\mu$W and 1.7 mW, respectively. The graphs in (c) and (d) are, respectively, the
calculated contrasts against Doppler background of probe transmittance for $\sigma_p^-$ and
$\sigma_p^+$ component with the parameters of $\Omega_p = \Omega_c = 2\gamma_a$. In all graphs, we
set $\delta_p = 0$. \label{fig2}}
\end{figure}

\begin{figure}
\includegraphics[height=8cm,width=3cm,angle=-90]{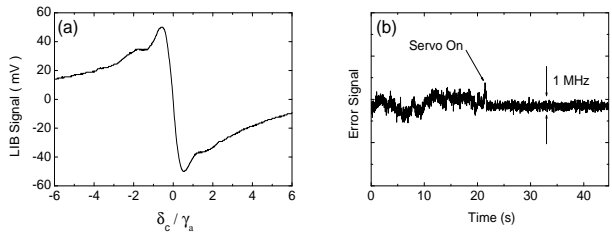}
\caption{(a) Doppler-free dispersive LIB signal measured by the SPD in Fig.~\ref{fig1}(b). (b)
Frequency fluctuation of the coupling laser before and after the servo-loop is on for frequency
stabilization. \label{fig3}}
\end{figure}
\end{document}